\documentclass[%
 reprint,
superscriptaddress,
 amsmath,amssymb,amsfonts,
 aps,
 nofootinbib,
]{revtex4-1}

\usepackage{graphicx}
\usepackage{dcolumn}
\usepackage{bm}

\newcommand{\adopt}{\alpha_d^{\rm opt}}

\begin{document}

\title{Biased thermodynamics can explain the behaviour of smart optimization algorithms that work above the dynamical threshold}

\author{Angelo Giorgio Cavaliere}
\email{angelog.cavaliere@gmail.com}
 \affiliation{
 Dipartimento di Fisica, Universit\`a ‘La Sapienza’, P.le A. Moro 5, 00185, Rome, Italy
}
\affiliation{Cybermedia Center, Osaka University, Toyonaka, Osaka,
560-0043, Japan}

\author{Federico Ricci-Tersenghi}%
\affiliation{
 Dipartimento di Fisica, Universit\`a ‘La Sapienza’, P.le A. Moro 5, 00185, Rome, Italy
}
\affiliation{
 INFN, Sezione di Roma1, and CNR-Nanotec, Rome unit, P.le A. Moro 5, 00185, Rome, Italy
}

\date{\today}

\begin{abstract}
Random constraint satisfaction problems can display a very rich structure in the space of solutions, with often an ergodicity breaking --- also known as clustering or dynamical --- transition preceding the satisfiability threshold when the constraint-to-variables ratio $\alpha$ is increased. However, smart algorithms start to fail finding solutions in polynomial time at some threshold $\alpha_{\rm alg}$ which is algorithmic dependent and generally bigger than the dynamical one $\alpha_d$. The reason for this discrepancy is due to the fact that $\alpha_d$ is traditionally computed according to the uniform measure over all the solutions. Thus, while bounding the region where a uniform sampling of the solutions is easy, it cannot predict the performance of off-equilibrium processes, that are still able of finding atypical solutions even beyond $\alpha_d$. Here we show that a reconciliation between algorithmic behaviour and thermodynamic prediction is nonetheless possible at least up to some threshold $\alpha_d^{\rm opt}\geq\alpha_d$, which is defined as the maximum value of the dynamical threshold computed on all possible probability measures over the solutions. We consider a simple Monte Carlo-based optimization algorithm, which is restricted to the solution space, and we demonstrate that sampling the equilibrium distribution of a biased measure improving on $\alpha_d$ is still possible even beyond the ergodicity breaking point for the uniform measure, where other algorithms hopelessly enter the out-of-equilibrium regime.
The conjecture we put forward is that many smart algorithms sample the solution space according to a biased measure: once this measure is identified, the algorithmic threshold is given by the corresponding ergodicity-breaking transition.
\end{abstract}

\maketitle


Many interesting physical processes are essentially out of equilibrium.
The first and more direct example is given by glassy models, which possess diverging relaxation timescales and live in out-of-equilibrium regimes for any experimental/observational time~\cite{bouchaudOutEquilibriumDynamics1997}.

A different, but even broader, class of processes that stay out of equilibrium are optimization or sampling algorithms that do not satisfy detail balance. The reason for this can be due either to the same definition of the algorithm, which is heuristic and does not satisfy any balance condition~\cite{braunsteinSurveyPropagationAlgorithm2006,seitzFocusedLocalSearch2005}, either because the simulation time is not large enough to achieve equilibrium~\cite{angeliniMonteCarloAlgorithms2019}.

Given that these interesting dynamical processes live in the off-equilibrium regime for most or all of the time, it is of primary importance to achieve an analytical description of this regime.
Unfortunately, this is very difficult and it has been achieved only in a very restricted class of models.
Essentially, the so-called dynamical mean-field equations can be written only for systems defined on a fully-connected topology, where couplings are required to scale as an inverse power of the system size~\cite{crisantiSphericalpspinInteractionSpinglass1993, cugliandoloAnalyticalSolutionOffequilibrium1993, cugliandoloOutofequilibriumRelaxationSherringtonKirkpatrick1994, bouchaudOutEquilibriumDynamics1997}. In other words, a close set of equations can be written only for models where the naive mean-field approximation holds, thanks to the couplings becoming very weak in the large N limit.

When considering more realistic systems where the couplings do not become small in the large $N$ limit, the situation is much more complex and some approximation is needed in order to try to provide a reasonable description of the out-of-equilibrium dynamics. Leaving apart finite-dimensional models, where any analytical treatment is out of question, it is worth considering models defined on sparse random graphs. On these graphs the finite degree of each variable allows couplings to remain finite in the large N limit, and nonetheless, the locally tree-like structure of the graph allows the use of the cavity method, which can provide the exact solution to the model in some regimes~\cite{mezardBetheLatticeSpin2001,mezardCavityMethodZero2003}.

The dynamical counterpart of the cavity method has been applied to several models defined on sparse random graphs to achieve an approximate description of the out-of-equilibrium dynamics~\cite{aurellCavityMasterEquation2017}. This approach is very promising but it suffers when a glass transition point is approached~\cite{aurellCavityMasterEquation2017,aurell_theory_2019}.

Among the open questions in the analytical description of these out-of-equilibrium processes, in particular optimization algorithms, there is the understanding of the limits of their performances. For example, in constraint satisfaction problems it is of primary importance to understand the  threshold experimented by algorithms searching for solutions and the reasons why above this threshold the search for solutions gets stuck.

Constraint satisfaction problems (CSP) are the prototype of optimization problems and one of the most studied problems in theoretical computer science~\cite{mezardInformationPhysicsComputation2012}.
In every CSP one has to search for an assignment of $N$ variables satisfying $M$ constraints.
More specifically, in random CSP, each of the $M$ constraints involves a randomly chosen small subset of variables (e.g.\ $K$ variables in a random $K$-satisfiability problem) and thus the interaction graph among variables is a sparse random graph if $M=\alpha N$, with the constraint-to-variables ratio $\alpha$ being constant.

Recent years have seen a large effort in using tools from statistical physics to understand the structure of the solutions space in random CSP~\cite{mezardRandomKsatisfiabilityProblem2002, krzakalaGibbsStatesSet2007, montanariClustersSolutionsReplica2008}.
From this line of research has emerged a very rich picture of the phase diagram changing the ratio $\alpha$. For $\alpha<\alpha_d$ the solutions are ``well connected'', in the sense that any dynamics changing $o(N)$ variables altogether can travel the whole solution space (in other words there is no ergodicity breaking among the majority of solutions). The value $\alpha_d$ is called a dynamical threshold because it corresponds to the breaking of ergodicity, above which solutions spontaneously form a clustered structure, which in turn does not allow local dynamics to sample the vast majority of solutions.

According to the above picture, we should expect any local algorithm\footnote{An algorithm changing $o(N)$ variables at each step.} to fail in the search for solutions if $\alpha>\alpha_d$. However, the numerical evidence which has been accumulated over the last few years tells a different story. Many smart, but heuristic, algorithms can find solutions above the dynamical threshold in a time growing at worst polynomially (and often linearly) with the system size $N$~\cite{seitzFocusedLocalSearch2005, ardeliusBehaviorHeuristicsLarge2006, krzakalaLandscapeAnalysisConstraint2007, dallastaEntropyLandscapeNonGibbs2008, braunsteinSurveyPropagationAlgorithm2006, marinoBacktrackingSurveyPropagation2016, angeliniMonteCarloAlgorithms2019}. How can these numerical observations be compatible with the dynamical transition taking place at $\alpha_d$? The best explanation at present is that smart algorithms actually do not sample solutions uniformly, while statistical physics computations were made using a uniform measure over the solutions.

A very clear and strong support to this explanation comes from a series of studies where solutions are sampled according to a non-uniform measure (while non-solutions are still assigned a zero weight)~\cite{krzakalaReweightedBeliefPropagation2012,braunsteinLargeDeviationsWhitening2016, baldassiSubdominantDenseClusters2015,  baldassiLocalEntropyMeasure2016, budzynskiBiasedLandscapesRandom2019, budzynskiBiasedMeasuresRandom2020, zhaoMaximallyFlexibleSolutions2020, maimbourgGeneratingDensePackings2018}. One of the main outcomes of these studies with \emph{biased measures} is that critical thresholds can change, including the dynamical one which is related to the appearance of barriers (both energetic and entropic). So it is clear that even without changing the set of solutions, a simple reweighting on this set can suppress entropic barriers, thus favouring the search for one of these solutions~\cite{bellittiEntropicBarriersReason2021}. One of the most emblematic cases supporting this line of thought is represented by the binary perceptron, for which has been recently proved~\cite{perkinsFrozen1RSBStructure2021,abbeProofContiguityConjecture2021} that typical (\emph{i.e.} almost all) solutions are completely frozen (isolated) in the clustered phase, and hence inaccessible to any known algorithm. However, efficient algorithms~\cite{braunsteinLearningMessagePassing2006,baldassiUnreasonableEffectivenessLearning2016}, are still capable of finding non-isolated solutions belonging to subdominant dense clusters~\cite{abbeBinaryPerceptronEfficient2021,baldassiUnveilingStructureWide2021}. These clusters of atypical solutions appear to connect solutions that may look otherwise as isolated due to entropic barriers~\cite{baldassiUnveilingStructureWide2021}.

A very appealing idea emerging from the above picture is the following. Let us concentrate on algorithms that sample the solutions space by moving between solutions (\emph{i.e.} they are restricted to the solution space) satisfying detailed balance according to any biased measure.
If a smart algorithm in this class does not sample solutions uniformly, but in a biased way, it may happen that when the uniform measure over solutions undergoes a clustering or dynamical phase transition, this algorithm is not affected and it can keep visiting the solution space without any ergodicity breaking until the dynamical threshold for the biased measure is achieved.

If the above idea is correct, we can then describe the large-time behaviour of such a smart algorithm by assuming it is sampling \emph{at equilibrium} the appropriate biased measure. Moreover, we can obtain the actual algorithmic threshold for this smart algorithm as the dynamical transition computed over the same biased measure.

In this work, we study a model corresponding to a CSP with continuous variables called the \emph{continuous coloring problem} and that undergoes a dynamical phase transition. Recently it has been shown how to optimize such a dynamical threshold by reweighting the space of solutions, via the modification of the interaction potential between pairs of variables. In the following we are going to present several numerical evidences that the best-performing algorithm searching for solutions by gradually increasing $\alpha$ does the following:
\begin{itemize}
    \item samples solutions according to the optimal biased measure computed in~\cite{cavaliereOptimizationDynamicTransition2021};
    \item is not affected by the dynamical transition happening at $\alpha_d$ in the uniform measure;
    \item remains at equilibrium before the dynamical transition $\adopt$, computed for the optimal biased measure maximising $\alpha_d$;
    \item the equilibration timescale diverges at $\adopt$, which is thus the algorithmic threshold for this smart algorithm.
\end{itemize}


\section{The model and the algorithm}
\subsection{The model}
In the continuous coloring $N$ real angular variables $x_i\in[0,2\pi)$, $i=1,\cdots,N$ are associated to the nodes of a sparse random graph and subjected to a constraint $\cos(x_i-x_j-\delta_{ij})\leq\cos\theta$ for each pair of vertices $(i,j)$ connected by an edge of the graph~\cite{cavaliereOptimizationDynamicTransition2021}. The parameter $\theta\in(0,\pi)$ is fixed and represents the minimum angular distance allowed between neighbours, while $\delta_{ij}\in[0,2\pi)$, $\forall (i,j)$ are uniform random shifts introduced in order to avoid having a periodic ordering of the variables. The model can be equivalently thought of as a XY spin-glass system~\cite{yoshinoDisorderfreeSpinGlass2018} or as an ideal glass of one-dimensional hard-spheres of diameter $\theta$, given the excluded volume nature of the interaction~\cite{mariJammingGlassTransitions2009,mezardSolutionSolvableModel2011}.
In this paper, we consider the behaviour of the model for typical instances extracted from the Erd\H{o}s-R\'enyi ensemble with average connectivity $2\alpha$, where $\alpha=M/N$ is the constraints-to-variables ratio. 

The phase diagram of the model when increasing $\alpha$ has been accurately obtained in~\cite{cavaliereOptimizationDynamicTransition2021}. For sufficiently small diameters $\theta$, the model undergoes a discontinuous glass transition belonging to the random first order (RFO) transition scheme. In particular, a dynamical or clustering transition is identified at $\alpha_d=34.63(2)$ for $\theta=\pi/10$, in a supporting model where the real angular variables are approximated with a high number ($p=200$) of clock-states. In this paper, we will stick to this value of $\theta$ and to the same discretization precision in order to compare the results from dynamics with the precise estimates for the transition thresholds. Notice that the quoted value of $\alpha_d$ refers to the location of the clustering or dynamical transition for the \emph{uniform} measure, which here corresponds to a purely hard-spheres interaction potential between neighbours on the graph.

Another outcome of~\cite{cavaliereOptimizationDynamicTransition2021} has been the computation of a biased interaction potential by specialising the approaches of~\cite{budzynskiBiasedLandscapesRandom2019} and~\cite{maimbourgGeneratingDensePackings2018}. This new biased measure is conveniently defined by a function of the interparticle angular distance $f(x)\equiv Z^{-1}\lim_{\beta\to\infty}e^{-\beta v(x)}$, where $v(x)$ is the pairwise energy potential (to be optimized), $\beta$ is the inverse of temperature and $Z$ an arbitrary normalization. The constraint satisfaction nature of the problem only requires $v(x)>0$ if the angular distance $x$ is smaller than $\theta$, so that $f(x)=0$. The uniform measure simply corresponds to the choice $f^{\rm flat}(x)\propto\Theta(x-\theta)$, where $\Theta(y)=0$ if $y<0$ and $\Theta(y)=1$ if $y\geq0$. A biased measure can instead show a non-trivial behaviour for $\theta\leq x\leq\pi$, while still respecting the constraints definition $f(x)=0$ for $0\leq x < \theta$.

In the following, we will define $f^{\rm opt}(x)$ as the optimal interaction function allowing one to postpone as much as possible to bigger $\alpha$'s the location of the dynamical transition. This has been empirically computed in~\cite{cavaliereOptimizationDynamicTransition2021}, allowing us to estimate $\adopt=37.71(1)$ as the maximum for $\alpha_d$ in the space of functions $f(x)$.
The resulting functional form for $f^{\rm opt}(x)$ corresponds to a short-range attraction for the spheres system, \emph{i.e.} tightly satisfied constraints are given more statistical weight than in the uniform measure, in analogy to what has been observed in the case of hard-spheres systems~\cite{dawsonHigherorderGlasstransitionSingularities2000,sciortinoOneLiquidTwo2002,maimbourgGeneratingDensePackings2018} (see the solid black line in figure~\ref{fig:2} below).

\begin{figure}[t]
    \centering
    \includegraphics[width=\columnwidth]{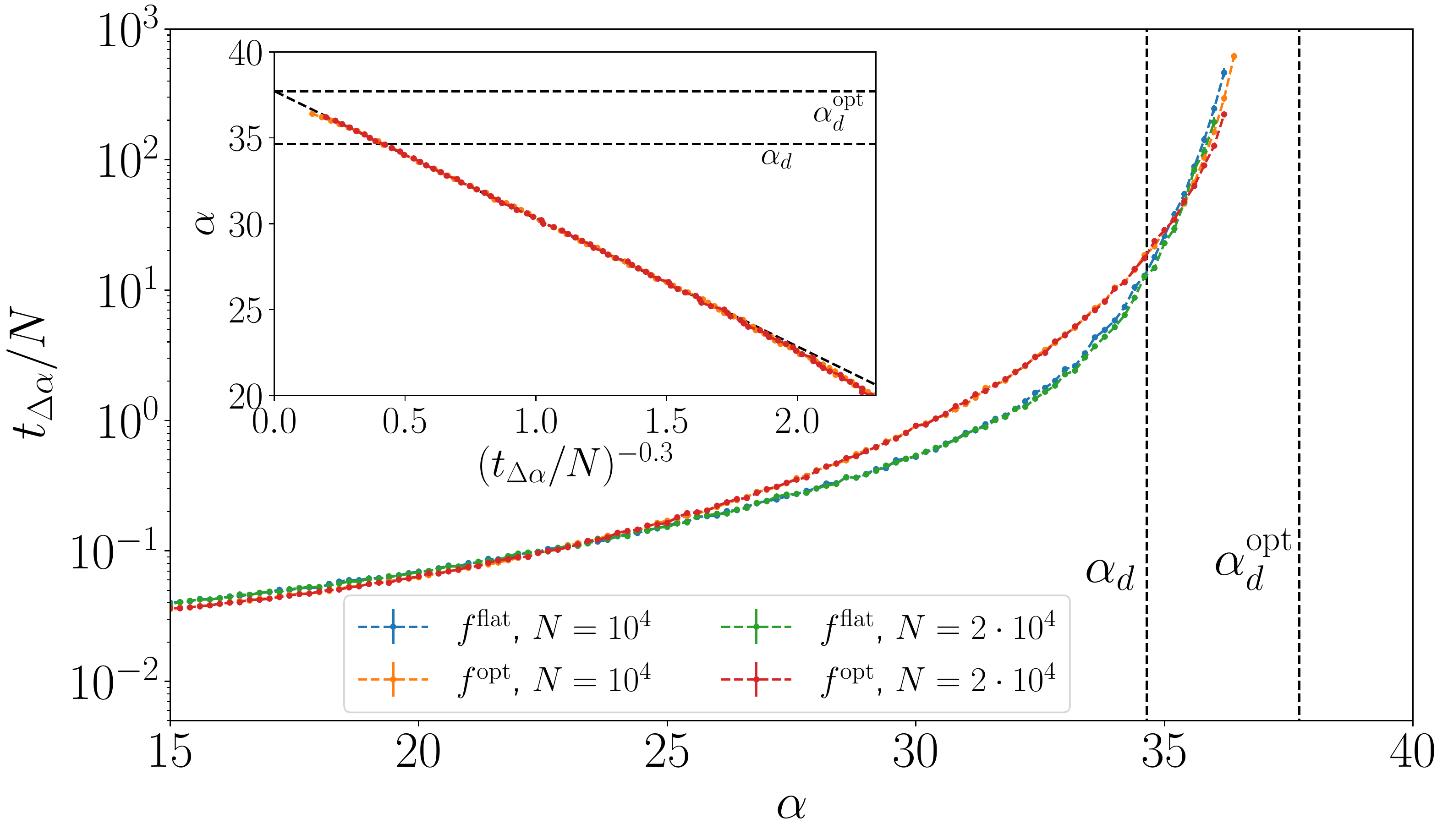}
    \caption{Time $t_{\Delta \alpha}$ needed in order to increase the number of links in the graph by $\Delta M = N\Delta \alpha$, with $\Delta\alpha=0.2$ a fixed smoothing parameter. Data is averaged over 55 (respectively 30) instances of the problem for $N=10^4$, $(2\cdot10^4)$. Both the uniform and optimized algorithms are able to find solutions in a time scaling linearly with the system size even beyond $\alpha_d$. The numerical procedure halts when the time to add a single link exceeds $10^7$ attempts. Inset: the optimized algorithm shows a smooth power-law divergence of the characteristic timescale at the predicted value $\adopt$ of the clustering transition for the optimally biased measure.}
    \label{fig:1}
\end{figure}

\subsection{Adiabatic protocol for constraint satisfaction}
We adapt to the continuous coloring problem an algorithm similar to the one presented in~\cite{krzakalaLandscapeAnalysisConstraint2007}, consisting in a gradual increase of the number of links in the graph up to a target density of constraints $\alpha_{\rm max}>\alpha_d$. The algorithm is characterized by the fact of being restricted to solution space at any time, and works as follows. Edges are initially stored in an ordered list (each one together with a uniform random shift) and removed from the graph, while angular variables are uniformly initialised. Then edges are proposed one at a time: if the relative constraint is satisfied that link is permanently added to the graph, otherwise a Monte Carlo sweep is performed to update the variables in the absence of the incriminated interaction, until that very constraint is satisfied and one can proceed to the following one. To implement the Monte Carlo, we adopt a heat bath rule according to $f^{\rm flat}(x)$ or $f^{\rm opt}(x)$ (notice that at each step the configuration of the variables is also a solution, so for each variable there always exists at least one state with non-zero probability). The update can be done efficiently thanks to the discretized nature of the variables.

This algorithm possesses two peculiar features, namely, it only moves between solutions and respects the detailed balance condition, and for both of these reasons we expect a direct connection between its behaviour and the thermodynamic description of the structure of the solutions space, \emph{i.e.} the location of $\alpha_d$. Despite its seemingly simple definition, our algorithm performs similarly to the very well-known simulated annealing algorithm. The reason for this efficiency is that, although it has no parameters to be optimized, it is in fact already obtained in a sort of adiabatic limit, since the algorithmic timescale naturally increases with the diverging of the relaxation time approaching $\adopt$, thus making it possible to stay at or very close to equilibrium (if the biased measure $f^{\rm opt}(x)$ is adopted along the procedure, as it will be shown in the following). 



\section{Results}
The same algorithm can run according to the uniform measure $f^{\rm flat}(x)$ or to the biased one $f^{\rm opt}(x)$. The general behaviour for both of them is depicted in Figure~\ref{fig:1}, where we plot on a log scale the time to add a fixed fraction of edges. There is almost no size dependence: thanks to the fact that we work on sparse random graphs, we can study very big systems as compared to the studies on fully-connected models, thus better approaching the large $N$ limit. 

A first comparison between the two processes shows that the algorithm running on $f^{\rm flat}$ performs better for intermediate values of $\alpha$, but seems to slow down the most in the long run (however this is not the principal result of our work, but rather it is the fact that for the algorithm running according to the biased measure we can provide analytic predictions). In the inset, we show our best extrapolation of what appears to be consistent with a power-law divergence of relaxation times for the protocol based on the optimized measure, and which is perfectly compatible with $\adopt$. It is worth noticing that also the algorithm adopting the uniform measure is able to easily surpass $\alpha_d$, and indeed we are going to show that it is going out of equilibrium (if it sampled the flat measure at equilibrium it should experience the ergodicity breaking happening at $\alpha_d$ and get stuck there).

\begin{figure}[t]
    \centering
    \includegraphics[width=\columnwidth]{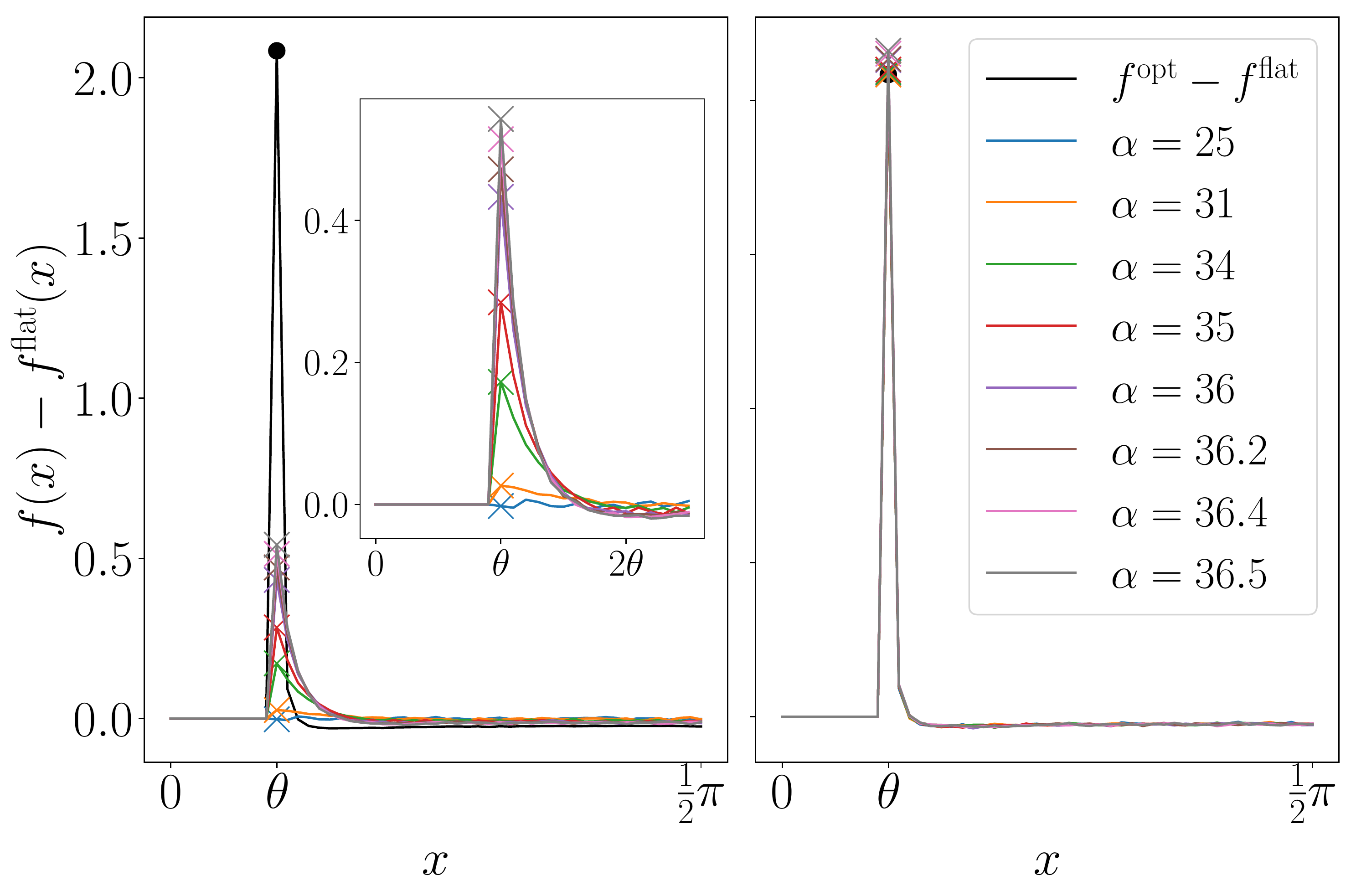}
    \caption{Distribution of angular distances (including shifts) between neighbours on the graph. We plot the difference between the measured distribution $f(x)$ and the function $f^{\rm flat}(x)\propto\Theta(x-\theta)$. All $f(x)$, $f^{\rm flat}(x)$ and $f^{\rm opt}(x)$ are arbitrarily normalized so that $\sum_{i=0}^{p-1}f(\frac{2\pi}{p}i)=p$, where $p=200$ is the number of clock states in our model. Different colors correspond to increasing values of $\alpha$ along the two MC protocols. Left panel: the algorithm using the uniform measure $f^{\rm flat}(x)$ spontaneously develops a peak for tightly satisfied constraints $x\approx\theta$ already before the ergodicity breaking threshold for the uniform measure $\alpha_d=34.63(2)$, and qualitatively resembling the shape of the distribution for the optimal measure. We believe this to be possibly linked to the good performance of the algorithm, which is clearly out-of-equilibrium since the observed non-trivial gaps distribution is different from the equilibrium expectation $f(x)=f^{\rm flat}(x)$, and also depending on $\alpha$. Right panel: using $f^{\rm opt}(x)$ the algorithm stays close to the equilibrium for the biased measure. Note that we have only reached values of $\alpha<\adopt$ for which the biased measure is still ergodic.}
    \label{fig:2}
\end{figure}

This is exemplified in Figure~\ref{fig:2}, where we argue that the algorithm running on $f^{\rm flat}$ is not sampling equilibrium, while the other one using $f^{\rm opt}$ is. To this end, we consider the distribution of angular distances 
(including shifts) between neighbours on the graph for different values of $\alpha$ along the run. We believe this to be a very significant physical observable, since dynamics strongly depends on the gaps actually present in the system, due to the excluded-volume nature of the problem. Moreover, whenever short loops in the network of interactions are absent, as also in the hard-spheres model in infinite dimensions~\cite{parisiTheorySimpleGlasses2020} or with infinitely ranged random shifts~\cite{mariDynamicalTransitionGlasses2011},
this pair correlation function is found at equilibrium (in the Replica Symmetric phase before the thermodynamic glass transition $\alpha_c$) to be simply proportional to the function $f(x)$ entering the definition of the measure. 

\begin{figure}[t]
    \centering
    \includegraphics[width=\columnwidth]{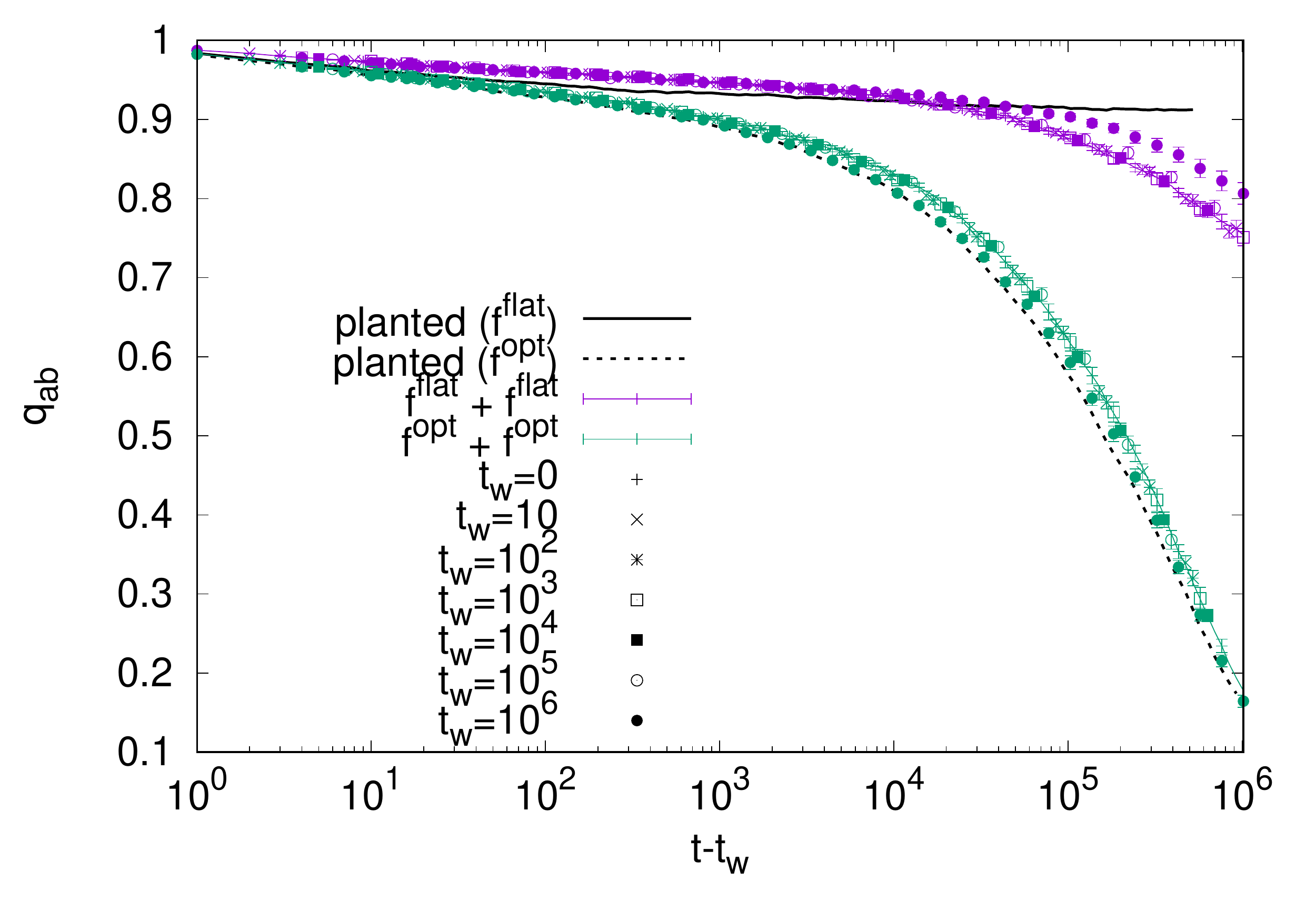}
    \caption{Relaxation Monte Carlo dynamics starting from the solutions found by our adiabatic search protocol at $\alpha=35.5>\alpha_d$. Here we simply use the same algorithm at fixed $\alpha=35.5$ and without changing the function $f$.
    After $t_w$ iterations, we duplicate the system and start measuring the overlap $q_{ab}$ between the two independently evolved replicas, as defined in~\cite{cavaliereOptimizationDynamicTransition2021}. The behaviour for solutions belonging to a typical cluster for the uniform measure (obtained through a planted initialization of the graph, which is valid here since we are before the condensation transition for this model, see~\cite{cavaliereOptimizationDynamicTransition2021}) is depicted by a black solid line. Solutions found by our optimization algorithm using $f^{\rm flat}$ (purple points) are very far from equilibrium behaviour: they do not belong to a proper cluster and start displaying aging for long waiting times. Solutions found by the optimized algorithm (green points) start on the contrary already very close to (and naturally move towards) the expected equilibrium behaviour for the biased measure (black dashed line).}
    \label{fig:3}
\end{figure}

Then from Figure~\ref{fig:2} we can observe how the algorithm using $f^{\rm flat}$ (left panel) is far from sampling the equilibrium according to the uniform measure, the pair correlation even becoming non-stationary before $\alpha_d$. But this is not the end of the story. Very interestingly, the pair correlation spontaneously evolves towards a distribution which is in some respect similar to the optimized one, in particular, it shows an increased number of tightly satisfied constraints for $x\approx\theta$: this is probably the reason why this algorithm manages to find solutions in linear time even beyond $\alpha_d$ and close to the optimized threshold $\adopt$. On the other hand, the behaviour of the optimized algorithm (right panel) is consistent with the equilibrium expectation for any $\alpha$ value.

We then carried out another experiment with the purpose of better characterizing the solutions found by the two algorithms. To this end, we performed a Monte Carlo exploration of the zero energy landscape of solutions starting from the solutions found by the two algorithms for fixed $\alpha=35.5$, a value which lies in the interesting region $\alpha_d<\alpha<\alpha_d^{\rm opt}$.

As shown in Figure~\ref{fig:3}, we observe once more that, in the attempt of continuing to sample solutions beyond $\alpha_d$, the uniform algorithm is forced to go out of equilibrium: its relaxation dynamics is different from the equilibrium one (black solid curve in the figure, obtained by initializing the graph via planting~\cite{krzakalaHidingQuietSolutions2009}) and for the biggest waiting times it displays a resumption of aging, as usually observed in systems living out of equilibrium. On the contrary, the dynamics of the algorithm with the optimized potential does not show any sign of aging and instead matches, after some time, the dynamics starting from an equilibrium (planted) solution for the biased measure.

We also checked, in a similar way as was done for example in~\cite{krzakalaReweightedBeliefPropagation2012}, that solutions planted above $\alpha_d$ inside a cluster for the uniform measure, are then able to decorrelate when evolved according to the optimized potential (still being always restricted to solution space). 

These findings are consistent with the idea that purely entropic barriers between clusters exist at least up to $\alpha_d^{\rm opt}$, and that one can use a smart biased measure in order to exploit rare paths and recover ergodicity beyond $\alpha_d$. Our approach also clearly suggests thinking about solutions found in the $\alpha>\alpha_d$ regime as the result of a strongly out-of-equilibrium procedure (like the protocol based on the uniform measure), that probably follows rare paths of solutions between clusters.
The aging of Figure~\ref{fig:3} is then interpreted as resulting from the difficulty for the system to find its way ``back'' inside a cluster (for the uniform measure), while being stuck in regions on the ``borders'' of such clusters, which are regions more likely to be selected by a biased measure that has not yet undergone a dynamical transition, as suggested from Figure~\ref{fig:2}.



\section{Discussion and perspectives}

In this paper, we have shown that it is possible to build physics-inspired algorithms for the optimization of random constraint satisfaction problems which are able to sample solutions at \emph{equilibrium} according to a properly biased measure also beyond the clustering transition for the uniform measure. This has been exemplified in the case of the continuous coloring, a model particularly interesting due to its connection to the problem of packing of one-dimensional spheres. 

As in other problems of hard-spheres, it was shown in~\cite{cavaliereOptimizationDynamicTransition2021} that adding a short-range attraction next to the hard-core repulsion one can extend the liquid phase of the system and postpone the transition point $\alpha_d$ to larger $\alpha$ values. This is a very physically-intuitive example of a \emph{biased} thermodynamics approach, which has become very popular in recent years also in the field of random constraint satisfaction problems and neural networks \cite{GabrieFarBeyond}. By exploiting rare paths of solutions, which would otherwise be entropically suppressed in the flat measure, the biased algorithm can recover ergodicity up to an analytically computable threshold $\alpha_d^{\rm opt}$, that performs remarkably well against simulations. 

An interesting point in considering the continuous coloring problem as we did in this work is that this model also allows for a real space pictorial interpretation of such rare paths of solutions. By appealing to the similitude with sticky spheres, we can figure out how the short-range attraction in the bias is helping the system to open void channels between particles, that in turn assist to recover the ergodicity since the interaction is of excluded-volume nature. This is clear for physical dynamics in two and three dimensions, but it is also true for one-dimensional particles if one considers a Monte Carlo dynamics, as we did, that allows jumping over neighbours, since in this case by closing contacts we can expect to make more room for particles to jump over their neighbours.
We believe this point of view provides a useful addition to the field of random constraint satisfaction problems, which usually do not allow for a real space intuitive representation.

Similar techniques to the ones illustrated in this work can be applied also to more standard CSP with discrete variables. For example, we plan to apply it to the hypergraph bicoloring problem, whose phase diagram is much richer, especially in presence of a bias~\cite{budzynskiBiasedLandscapesRandom2019}.
It is worth stressing that recent work on planted random graph coloring~\cite{angeliniLimits2022} has found a tight connection between the algorithmic threshold for Monte Carlo-based algorithms and the thermodynamic phase transition in a modified model where several replicas are coupled. Given that the coupling among replicas is in some sense equivalent to reweighting solutions according to the local entropy~\cite{baldassiLocalEntropyMeasure2016}, this is another example where algorithmic thresholds are connected to  phase transitions in a biased measure.

\begin{acknowledgments}
This research has been supported by ICSC Centro
Nazionale di Ricerca in High Performance Computing, Big
Data and Quantum Computing, funded by European Union
NextGenerationEU.
\end{acknowledgments}

\bibliography{main}

\providecommand{\noopsort}[1]{}\providecommand{\singleletter}[1]{#1}%
\begin{thebibliography}{45}%
\makeatletter
\providecommand \@ifxundefined [1]{%
 \@ifx{#1\undefined}
}%
\providecommand \@ifnum [1]{%
 \ifnum #1\expandafter \@firstoftwo
 \else \expandafter \@secondoftwo
 \fi
}%
\providecommand \@ifx [1]{%
 \ifx #1\expandafter \@firstoftwo
 \else \expandafter \@secondoftwo
 \fi
}%
\providecommand \natexlab [1]{#1}%
\providecommand \enquote  [1]{``#1''}%
\providecommand \bibnamefont  [1]{#1}%
\providecommand \bibfnamefont [1]{#1}%
\providecommand \citenamefont [1]{#1}%
\providecommand \href@noop [0]{\@secondoftwo}%
\providecommand \href [0]{\begingroup \@sanitize@url \@href}%
\providecommand \@href[1]{\@@startlink{#1}\@@href}%
\providecommand \@@href[1]{\endgroup#1\@@endlink}%
\providecommand \@sanitize@url [0]{\catcode `\\12\catcode `\$12\catcode
  `\&12\catcode `\#12\catcode `\^12\catcode `\_12\catcode `\%12\relax}%
\providecommand \@@startlink[1]{}%
\providecommand \@@endlink[0]{}%
\providecommand \url  [0]{\begingroup\@sanitize@url \@url }%
\providecommand \@url [1]{\endgroup\@href {#1}{\urlprefix }}%
\providecommand \urlprefix  [0]{URL }%
\providecommand \Eprint [0]{\href }%
\providecommand \doibase [0]{http://dx.doi.org/}%
\providecommand \selectlanguage [0]{\@gobble}%
\providecommand \bibinfo  [0]{\@secondoftwo}%
\providecommand \bibfield  [0]{\@secondoftwo}%
\providecommand \translation [1]{[#1]}%
\providecommand \BibitemOpen [0]{}%
\providecommand \bibitemStop [0]{}%
\providecommand \bibitemNoStop [0]{.\EOS\space}%
\providecommand \EOS [0]{\spacefactor3000\relax}%
\providecommand \BibitemShut  [1]{\csname bibitem#1\endcsname}%
\let\auto@bib@innerbib\@empty
\bibitem [{\citenamefont {Bouchaud}\ \emph {et~al.}(1997)\citenamefont
  {Bouchaud}, \citenamefont {Cugliandolo}, \citenamefont {Kurchan},\ and\
  \citenamefont {M{\'e}zard}}]{bouchaudOutEquilibriumDynamics1997}%
  \BibitemOpen
  \bibfield  {author} {\bibinfo {author} {\bibfnamefont {J.-P.}\ \bibnamefont
  {Bouchaud}}, \bibinfo {author} {\bibfnamefont {L.~F.}\ \bibnamefont
  {Cugliandolo}}, \bibinfo {author} {\bibfnamefont {J.}~\bibnamefont
  {Kurchan}}, \ and\ \bibinfo {author} {\bibfnamefont {M.}~\bibnamefont
  {M{\'e}zard}},\ }in\ \href {\doibase 10.1142/9789812819437_0006} {\emph
  {\bibinfo {booktitle} {Spin {{Glasses}} and {{Random Fields}}}}},\ \bibinfo
  {series} {Series on {{Directions}} in {{Condensed Matter Physics}}},
  Vol.~\bibinfo {volume} {12}\ (\bibinfo  {publisher} {{WORLD SCIENTIFIC}},\
  \bibinfo {year} {1997})\ pp.\ \bibinfo {pages} {161--223},\ \Eprint
  {http://arxiv.org/abs/cond-mat/9702070} {arXiv:cond-mat/9702070} \BibitemShut
  {NoStop}%
\bibitem [{\citenamefont {Braunstein}\ \emph {et~al.}(2006)\citenamefont
  {Braunstein}, \citenamefont {Mezard},\ and\ \citenamefont
  {Zecchina}}]{braunsteinSurveyPropagationAlgorithm2006}%
  \BibitemOpen
  \bibfield  {author} {\bibinfo {author} {\bibfnamefont {A.}~\bibnamefont
  {Braunstein}}, \bibinfo {author} {\bibfnamefont {M.}~\bibnamefont {Mezard}},
  \ and\ \bibinfo {author} {\bibfnamefont {R.}~\bibnamefont {Zecchina}},\
  }\href@noop {} {\bibfield  {journal} {\bibinfo  {journal} {arXiv:cs/0212002}\
  } (\bibinfo {year} {2006})}\BibitemShut {NoStop}%
\bibitem [{\citenamefont {Seitz}\ \emph {et~al.}(2005)\citenamefont {Seitz},
  \citenamefont {Alava},\ and\ \citenamefont
  {Orponen}}]{seitzFocusedLocalSearch2005}%
  \BibitemOpen
  \bibfield  {author} {\bibinfo {author} {\bibfnamefont {S.}~\bibnamefont
  {Seitz}}, \bibinfo {author} {\bibfnamefont {M.}~\bibnamefont {Alava}}, \ and\
  \bibinfo {author} {\bibfnamefont {P.}~\bibnamefont {Orponen}},\ }\href
  {\doibase 10.1088/1742-5468/2005/06/P06006} {\bibfield  {journal} {\bibinfo
  {journal} {Journal of Statistical Mechanics: Theory and Experiment}\ }\textbf
  {\bibinfo {volume} {2005}},\ \bibinfo {pages} {P06006} (\bibinfo {year}
  {2005})}\BibitemShut {NoStop}%
\bibitem [{\citenamefont {Angelini}\ and\ \citenamefont
  {{Ricci-Tersenghi}}(2019)}]{angeliniMonteCarloAlgorithms2019}%
  \BibitemOpen
  \bibfield  {author} {\bibinfo {author} {\bibfnamefont {M.~C.}\ \bibnamefont
  {Angelini}}\ and\ \bibinfo {author} {\bibfnamefont {F.}~\bibnamefont
  {{Ricci-Tersenghi}}},\ }\href {\doibase 10.1103/PhysRevE.100.013302}
  {\bibfield  {journal} {\bibinfo  {journal} {Physical Review E}\ }\textbf
  {\bibinfo {volume} {100}},\ \bibinfo {pages} {013302} (\bibinfo {year}
  {2019})}\BibitemShut {NoStop}%
\bibitem [{\citenamefont {Crisanti}\ \emph {et~al.}(1993)\citenamefont
  {Crisanti}, \citenamefont {Horner},\ and\ \citenamefont
  {Sommers}}]{crisantiSphericalpspinInteractionSpinglass1993}%
  \BibitemOpen
  \bibfield  {author} {\bibinfo {author} {\bibfnamefont {A.}~\bibnamefont
  {Crisanti}}, \bibinfo {author} {\bibfnamefont {H.}~\bibnamefont {Horner}}, \
  and\ \bibinfo {author} {\bibfnamefont {H.~J.}\ \bibnamefont {Sommers}},\
  }\href {\doibase 10.1007/BF01312184} {\bibfield  {journal} {\bibinfo
  {journal} {Zeitschrift f\"ur Physik B Condensed Matter}\ }\textbf {\bibinfo
  {volume} {92}},\ \bibinfo {pages} {257} (\bibinfo {year} {1993})}\BibitemShut
  {NoStop}%
\bibitem [{\citenamefont {Cugliandolo}\ and\ \citenamefont
  {Kurchan}(1993)}]{cugliandoloAnalyticalSolutionOffequilibrium1993}%
  \BibitemOpen
  \bibfield  {author} {\bibinfo {author} {\bibfnamefont {L.~F.}\ \bibnamefont
  {Cugliandolo}}\ and\ \bibinfo {author} {\bibfnamefont {J.}~\bibnamefont
  {Kurchan}},\ }\href {\doibase 10.1103/PhysRevLett.71.173} {\bibfield
  {journal} {\bibinfo  {journal} {Physical Review Letters}\ }\textbf {\bibinfo
  {volume} {71}},\ \bibinfo {pages} {173} (\bibinfo {year} {1993})}\BibitemShut
  {NoStop}%
\bibitem [{\citenamefont {Cugliandolo}\ and\ \citenamefont
  {Kurchan}(1994)}]{cugliandoloOutofequilibriumRelaxationSherringtonKirkpatrick1994}%
  \BibitemOpen
  \bibfield  {author} {\bibinfo {author} {\bibfnamefont {L.~F.}\ \bibnamefont
  {Cugliandolo}}\ and\ \bibinfo {author} {\bibfnamefont {J.}~\bibnamefont
  {Kurchan}},\ }\href {\doibase 10.1088/0305-4470/27/17/011} {\bibfield
  {journal} {\bibinfo  {journal} {Journal of Physics A: Mathematical and
  General}\ }\textbf {\bibinfo {volume} {27}},\ \bibinfo {pages} {5749}
  (\bibinfo {year} {1994})}\BibitemShut {NoStop}%
\bibitem [{\citenamefont {Mezard}\ and\ \citenamefont
  {Parisi}(2001)}]{mezardBetheLatticeSpin2001}%
  \BibitemOpen
  \bibfield  {author} {\bibinfo {author} {\bibfnamefont {M.}~\bibnamefont
  {Mezard}}\ and\ \bibinfo {author} {\bibfnamefont {G.}~\bibnamefont
  {Parisi}},\ }\href {\doibase 10.1007/PL00011099} {\bibfield  {journal}
  {\bibinfo  {journal} {The European Physical Journal B}\ }\textbf {\bibinfo
  {volume} {20}},\ \bibinfo {pages} {217} (\bibinfo {year} {2001})},\ \Eprint
  {http://arxiv.org/abs/cond-mat/0009418} {arXiv:cond-mat/0009418} \BibitemShut
  {NoStop}%
\bibitem [{\citenamefont {M{\'e}zard}\ and\ \citenamefont
  {Parisi}(2003)}]{mezardCavityMethodZero2003}%
  \BibitemOpen
  \bibfield  {author} {\bibinfo {author} {\bibfnamefont {M.}~\bibnamefont
  {M{\'e}zard}}\ and\ \bibinfo {author} {\bibfnamefont {G.}~\bibnamefont
  {Parisi}},\ }\href {\doibase 10.1023/A:1022221005097} {\bibfield  {journal}
  {\bibinfo  {journal} {Journal of Statistical Physics}\ }\textbf {\bibinfo
  {volume} {111}},\ \bibinfo {pages} {1} (\bibinfo {year} {2003})}\BibitemShut
  {NoStop}%
\bibitem [{\citenamefont {Aurell}\ \emph {et~al.}(2017)\citenamefont {Aurell},
  \citenamefont {Del~Ferraro}, \citenamefont {Dom{\'i}nguez},\ and\
  \citenamefont {Mulet}}]{aurellCavityMasterEquation2017}%
  \BibitemOpen
  \bibfield  {author} {\bibinfo {author} {\bibfnamefont {E.}~\bibnamefont
  {Aurell}}, \bibinfo {author} {\bibfnamefont {G.}~\bibnamefont {Del~Ferraro}},
  \bibinfo {author} {\bibfnamefont {E.}~\bibnamefont {Dom{\'i}nguez}}, \ and\
  \bibinfo {author} {\bibfnamefont {R.}~\bibnamefont {Mulet}},\ }\href
  {\doibase 10.1103/PhysRevE.95.052119} {\bibfield  {journal} {\bibinfo
  {journal} {Physical Review E}\ }\textbf {\bibinfo {volume} {95}},\ \bibinfo
  {pages} {052119} (\bibinfo {year} {2017})}\BibitemShut {NoStop}%
\bibitem [{\citenamefont {Aurell}\ \emph {et~al.}(2019)\citenamefont {Aurell},
  \citenamefont {Domínguez}, \citenamefont {Machado},\ and\ \citenamefont
  {Mulet}}]{aurell_theory_2019}%
  \BibitemOpen
  \bibfield  {author} {\bibinfo {author} {\bibfnamefont {E.}~\bibnamefont
  {Aurell}}, \bibinfo {author} {\bibfnamefont {E.}~\bibnamefont {Domínguez}},
  \bibinfo {author} {\bibfnamefont {D.}~\bibnamefont {Machado}}, \ and\
  \bibinfo {author} {\bibfnamefont {R.}~\bibnamefont {Mulet}},\ }\href
  {\doibase 10.1103/PhysRevLett.123.230602} {\bibfield  {journal} {\bibinfo
  {journal} {Physical Review Letters}\ }\textbf {\bibinfo {volume} {123}},\
  \bibinfo {pages} {230602} (\bibinfo {year} {2019})},\ \bibinfo {note}
  {publisher: American Physical Society}\BibitemShut {NoStop}%
\bibitem [{\citenamefont {M{\'e}zard}\ and\ \citenamefont
  {Montanari}(2012)}]{mezardInformationPhysicsComputation2012}%
  \BibitemOpen
  \bibfield  {author} {\bibinfo {author} {\bibfnamefont {M.}~\bibnamefont
  {M{\'e}zard}}\ and\ \bibinfo {author} {\bibfnamefont {A.}~\bibnamefont
  {Montanari}},\ }\href@noop {} {\emph {\bibinfo {title} {Information, Physics,
  and Computation}}},\ Oxford Graduate Texts\ (\bibinfo  {publisher} {{Oxford
  Univ. Press}},\ \bibinfo {address} {{Oxford}},\ \bibinfo {year}
  {2012})\BibitemShut {NoStop}%
\bibitem [{\citenamefont {M{\'e}zard}\ and\ \citenamefont
  {Zecchina}(2002)}]{mezardRandomKsatisfiabilityProblem2002}%
  \BibitemOpen
  \bibfield  {author} {\bibinfo {author} {\bibfnamefont {M.}~\bibnamefont
  {M{\'e}zard}}\ and\ \bibinfo {author} {\bibfnamefont {R.}~\bibnamefont
  {Zecchina}},\ }\href {\doibase 10.1103/PhysRevE.66.056126} {\bibfield
  {journal} {\bibinfo  {journal} {Physical Review E}\ }\textbf {\bibinfo
  {volume} {66}},\ \bibinfo {pages} {056126} (\bibinfo {year}
  {2002})}\BibitemShut {NoStop}%
\bibitem [{\citenamefont {Krzakala}\ \emph {et~al.}(2007)\citenamefont
  {Krzakala}, \citenamefont {Montanari}, \citenamefont {{Ricci-Tersenghi}},
  \citenamefont {Semerjian},\ and\ \citenamefont
  {Zdeborova}}]{krzakalaGibbsStatesSet2007}%
  \BibitemOpen
  \bibfield  {author} {\bibinfo {author} {\bibfnamefont {F.}~\bibnamefont
  {Krzakala}}, \bibinfo {author} {\bibfnamefont {A.}~\bibnamefont {Montanari}},
  \bibinfo {author} {\bibfnamefont {F.}~\bibnamefont {{Ricci-Tersenghi}}},
  \bibinfo {author} {\bibfnamefont {G.}~\bibnamefont {Semerjian}}, \ and\
  \bibinfo {author} {\bibfnamefont {L.}~\bibnamefont {Zdeborova}},\ }\href
  {\doibase 10.1073/pnas.0703685104} {\bibfield  {journal} {\bibinfo  {journal}
  {Proceedings of the National Academy of Sciences}\ }\textbf {\bibinfo
  {volume} {104}},\ \bibinfo {pages} {10318} (\bibinfo {year}
  {2007})}\BibitemShut {NoStop}%
\bibitem [{\citenamefont {Montanari}\ \emph {et~al.}(2008)\citenamefont
  {Montanari}, \citenamefont {{Ricci-Tersenghi}},\ and\ \citenamefont
  {Semerjian}}]{montanariClustersSolutionsReplica2008}%
  \BibitemOpen
  \bibfield  {author} {\bibinfo {author} {\bibfnamefont {A.}~\bibnamefont
  {Montanari}}, \bibinfo {author} {\bibfnamefont {F.}~\bibnamefont
  {{Ricci-Tersenghi}}}, \ and\ \bibinfo {author} {\bibfnamefont
  {G.}~\bibnamefont {Semerjian}},\ }\href {\doibase
  10.1088/1742-5468/2008/04/P04004} {\bibfield  {journal} {\bibinfo  {journal}
  {Journal of Statistical Mechanics: Theory and Experiment}\ }\textbf {\bibinfo
  {volume} {2008}},\ \bibinfo {pages} {P04004} (\bibinfo {year}
  {2008})}\BibitemShut {NoStop}%
\bibitem [{\citenamefont {Ardelius}\ and\ \citenamefont
  {Aurell}(2006)}]{ardeliusBehaviorHeuristicsLarge2006}%
  \BibitemOpen
  \bibfield  {author} {\bibinfo {author} {\bibfnamefont {J.}~\bibnamefont
  {Ardelius}}\ and\ \bibinfo {author} {\bibfnamefont {E.}~\bibnamefont
  {Aurell}},\ }\href {\doibase 10.1103/PhysRevE.74.037702} {\bibfield
  {journal} {\bibinfo  {journal} {Physical Review E}\ }\textbf {\bibinfo
  {volume} {74}},\ \bibinfo {pages} {037702} (\bibinfo {year}
  {2006})}\BibitemShut {NoStop}%
\bibitem [{\citenamefont {Krzakala}\ and\ \citenamefont
  {Kurchan}(2007)}]{krzakalaLandscapeAnalysisConstraint2007}%
  \BibitemOpen
  \bibfield  {author} {\bibinfo {author} {\bibfnamefont {F.}~\bibnamefont
  {Krzakala}}\ and\ \bibinfo {author} {\bibfnamefont {J.}~\bibnamefont
  {Kurchan}},\ }\href {\doibase 10.1103/PhysRevE.76.021122} {\bibfield
  {journal} {\bibinfo  {journal} {Physical Review E}\ }\textbf {\bibinfo
  {volume} {76}},\ \bibinfo {pages} {021122} (\bibinfo {year}
  {2007})}\BibitemShut {NoStop}%
\bibitem [{\citenamefont {Dall'Asta}\ \emph {et~al.}(2008)\citenamefont
  {Dall'Asta}, \citenamefont {Ramezanpour},\ and\ \citenamefont
  {Zecchina}}]{dallastaEntropyLandscapeNonGibbs2008}%
  \BibitemOpen
  \bibfield  {author} {\bibinfo {author} {\bibfnamefont {L.}~\bibnamefont
  {Dall'Asta}}, \bibinfo {author} {\bibfnamefont {A.}~\bibnamefont
  {Ramezanpour}}, \ and\ \bibinfo {author} {\bibfnamefont {R.}~\bibnamefont
  {Zecchina}},\ }\href {\doibase 10.1103/PhysRevE.77.031118} {\bibfield
  {journal} {\bibinfo  {journal} {Physical Review E}\ }\textbf {\bibinfo
  {volume} {77}},\ \bibinfo {pages} {031118} (\bibinfo {year}
  {2008})}\BibitemShut {NoStop}%
\bibitem [{\citenamefont {Marino}\ \emph {et~al.}(2016)\citenamefont {Marino},
  \citenamefont {Parisi},\ and\ \citenamefont
  {{Ricci-Tersenghi}}}]{marinoBacktrackingSurveyPropagation2016}%
  \BibitemOpen
  \bibfield  {author} {\bibinfo {author} {\bibfnamefont {R.}~\bibnamefont
  {Marino}}, \bibinfo {author} {\bibfnamefont {G.}~\bibnamefont {Parisi}}, \
  and\ \bibinfo {author} {\bibfnamefont {F.}~\bibnamefont
  {{Ricci-Tersenghi}}},\ }\href {\doibase 10.1038/ncomms12996} {\bibfield
  {journal} {\bibinfo  {journal} {Nature Communications}\ }\textbf {\bibinfo
  {volume} {7}},\ \bibinfo {pages} {12996} (\bibinfo {year}
  {2016})}\BibitemShut {NoStop}%
\bibitem [{\citenamefont {Krzakala}\ \emph {et~al.}(2012)\citenamefont
  {Krzakala}, \citenamefont {M{\'e}zard},\ and\ \citenamefont
  {Zdeborov{\'a}}}]{krzakalaReweightedBeliefPropagation2012}%
  \BibitemOpen
  \bibfield  {author} {\bibinfo {author} {\bibfnamefont {F.}~\bibnamefont
  {Krzakala}}, \bibinfo {author} {\bibfnamefont {M.}~\bibnamefont
  {M{\'e}zard}}, \ and\ \bibinfo {author} {\bibfnamefont {L.}~\bibnamefont
  {Zdeborov{\'a}}},\ }\href {\doibase 10.3233/SAT190096} {\bibfield  {journal}
  {\bibinfo  {journal} {Journal on Satisfiability, Boolean Modeling and
  Computation}\ }\textbf {\bibinfo {volume} {8}},\ \bibinfo {pages} {149}
  (\bibinfo {year} {2012})}\BibitemShut {NoStop}%
\bibitem [{\citenamefont {Braunstein}\ \emph {et~al.}(2016)\citenamefont
  {Braunstein}, \citenamefont {Dall'Asta}, \citenamefont {Semerjian},\ and\
  \citenamefont {Zdeborov{\'a}}}]{braunsteinLargeDeviationsWhitening2016}%
  \BibitemOpen
  \bibfield  {author} {\bibinfo {author} {\bibfnamefont {A.}~\bibnamefont
  {Braunstein}}, \bibinfo {author} {\bibfnamefont {L.}~\bibnamefont
  {Dall'Asta}}, \bibinfo {author} {\bibfnamefont {G.}~\bibnamefont
  {Semerjian}}, \ and\ \bibinfo {author} {\bibfnamefont {L.}~\bibnamefont
  {Zdeborov{\'a}}},\ }\href {\doibase 10.1088/1742-5468/2016/05/053401}
  {\bibfield  {journal} {\bibinfo  {journal} {Journal of Statistical Mechanics:
  Theory and Experiment}\ }\textbf {\bibinfo {volume} {2016}},\ \bibinfo
  {pages} {053401} (\bibinfo {year} {2016})}\BibitemShut {NoStop}%
\bibitem [{\citenamefont {Baldassi}\ \emph {et~al.}(2015)\citenamefont
  {Baldassi}, \citenamefont {Ingrosso}, \citenamefont {Lucibello},
  \citenamefont {Saglietti},\ and\ \citenamefont
  {Zecchina}}]{baldassiSubdominantDenseClusters2015}%
  \BibitemOpen
  \bibfield  {author} {\bibinfo {author} {\bibfnamefont {C.}~\bibnamefont
  {Baldassi}}, \bibinfo {author} {\bibfnamefont {A.}~\bibnamefont {Ingrosso}},
  \bibinfo {author} {\bibfnamefont {C.}~\bibnamefont {Lucibello}}, \bibinfo
  {author} {\bibfnamefont {L.}~\bibnamefont {Saglietti}}, \ and\ \bibinfo
  {author} {\bibfnamefont {R.}~\bibnamefont {Zecchina}},\ }\href {\doibase
  10.1103/PhysRevLett.115.128101} {\bibfield  {journal} {\bibinfo  {journal}
  {Physical Review Letters}\ }\textbf {\bibinfo {volume} {115}},\ \bibinfo
  {pages} {128101} (\bibinfo {year} {2015})}\BibitemShut {NoStop}%
\bibitem [{\citenamefont {Baldassi}\ \emph
  {et~al.}(2016{\natexlab{a}})\citenamefont {Baldassi}, \citenamefont
  {Ingrosso}, \citenamefont {Lucibello}, \citenamefont {Saglietti},\ and\
  \citenamefont {Zecchina}}]{baldassiLocalEntropyMeasure2016}%
  \BibitemOpen
  \bibfield  {author} {\bibinfo {author} {\bibfnamefont {C.}~\bibnamefont
  {Baldassi}}, \bibinfo {author} {\bibfnamefont {A.}~\bibnamefont {Ingrosso}},
  \bibinfo {author} {\bibfnamefont {C.}~\bibnamefont {Lucibello}}, \bibinfo
  {author} {\bibfnamefont {L.}~\bibnamefont {Saglietti}}, \ and\ \bibinfo
  {author} {\bibfnamefont {R.}~\bibnamefont {Zecchina}},\ }\href {\doibase
  10.1088/1742-5468/2016/02/023301} {\bibfield  {journal} {\bibinfo  {journal}
  {Journal of Statistical Mechanics: Theory and Experiment}\ }\textbf {\bibinfo
  {volume} {2016}},\ \bibinfo {pages} {023301} (\bibinfo {year}
  {2016}{\natexlab{a}})},\ \Eprint {http://arxiv.org/abs/1511.05634}
  {arXiv:1511.05634} \BibitemShut {NoStop}%
\bibitem [{\citenamefont {Budzynski}\ \emph {et~al.}(2019)\citenamefont
  {Budzynski}, \citenamefont {{Ricci-Tersenghi}},\ and\ \citenamefont {{Guilhem
  Semerjian}}}]{budzynskiBiasedLandscapesRandom2019}%
  \BibitemOpen
  \bibfield  {author} {\bibinfo {author} {\bibfnamefont {L.}~\bibnamefont
  {Budzynski}}, \bibinfo {author} {\bibfnamefont {F.}~\bibnamefont
  {{Ricci-Tersenghi}}}, \ and\ \bibinfo {author} {\bibnamefont {{Guilhem
  Semerjian}}},\ }\href {\doibase 10.1088/1742-5468/ab02de} {\bibfield
  {journal} {\bibinfo  {journal} {Journal of Statistical Mechanics: Theory and
  Experiment}\ }\textbf {\bibinfo {volume} {2019}},\ \bibinfo {pages} {023302}
  (\bibinfo {year} {2019})}\BibitemShut {NoStop}%
\bibitem [{\citenamefont {Budzynski}\ and\ \citenamefont
  {Semerjian}(2020)}]{budzynskiBiasedMeasuresRandom2020}%
  \BibitemOpen
  \bibfield  {author} {\bibinfo {author} {\bibfnamefont {L.}~\bibnamefont
  {Budzynski}}\ and\ \bibinfo {author} {\bibfnamefont {G.}~\bibnamefont
  {Semerjian}},\ }\href {\doibase 10.1088/1742-5468/abb8c8} {\bibfield
  {journal} {\bibinfo  {journal} {Journal of Statistical Mechanics: Theory and
  Experiment}\ }\textbf {\bibinfo {volume} {2020}},\ \bibinfo {pages} {103406}
  (\bibinfo {year} {2020})}\BibitemShut {NoStop}%
\bibitem [{\citenamefont {Zhao}\ and\ \citenamefont
  {Zhou}(2020)}]{zhaoMaximallyFlexibleSolutions2020}%
  \BibitemOpen
  \bibfield  {author} {\bibinfo {author} {\bibfnamefont {H.}~\bibnamefont
  {Zhao}}\ and\ \bibinfo {author} {\bibfnamefont {H.-J.}\ \bibnamefont
  {Zhou}},\ }\href {\doibase 10.1103/PhysRevE.102.012301} {\bibfield  {journal}
  {\bibinfo  {journal} {Physical Review E}\ }\textbf {\bibinfo {volume}
  {102}},\ \bibinfo {pages} {012301} (\bibinfo {year} {2020})}\BibitemShut
  {NoStop}%
\bibitem [{\citenamefont {Maimbourg}\ \emph {et~al.}(2018)\citenamefont
  {Maimbourg}, \citenamefont {Sellitto}, \citenamefont {Semerjian},\ and\
  \citenamefont {Zamponi}}]{maimbourgGeneratingDensePackings2018}%
  \BibitemOpen
  \bibfield  {author} {\bibinfo {author} {\bibfnamefont {T.}~\bibnamefont
  {Maimbourg}}, \bibinfo {author} {\bibfnamefont {M.}~\bibnamefont {Sellitto}},
  \bibinfo {author} {\bibfnamefont {G.}~\bibnamefont {Semerjian}}, \ and\
  \bibinfo {author} {\bibfnamefont {F.}~\bibnamefont {Zamponi}},\ }\href
  {\doibase 10.21468/SciPostPhys.4.6.039} {\bibfield  {journal} {\bibinfo
  {journal} {SciPost Physics}\ }\textbf {\bibinfo {volume} {4}},\ \bibinfo
  {pages} {039} (\bibinfo {year} {2018})}\BibitemShut {NoStop}%
\bibitem [{\citenamefont {Bellitti}\ \emph {et~al.}(2021)\citenamefont
  {Bellitti}, \citenamefont {{Ricci-Tersenghi}},\ and\ \citenamefont
  {Scardicchio}}]{bellittiEntropicBarriersReason2021}%
  \BibitemOpen
  \bibfield  {author} {\bibinfo {author} {\bibfnamefont {M.}~\bibnamefont
  {Bellitti}}, \bibinfo {author} {\bibfnamefont {F.}~\bibnamefont
  {{Ricci-Tersenghi}}}, \ and\ \bibinfo {author} {\bibfnamefont
  {A.}~\bibnamefont {Scardicchio}},\ }\href {\doibase
  10.1103/PhysRevResearch.3.043015} {\bibfield  {journal} {\bibinfo  {journal}
  {Physical Review Research}\ }\textbf {\bibinfo {volume} {3}},\ \bibinfo
  {pages} {043015} (\bibinfo {year} {2021})}\BibitemShut {NoStop}%
\bibitem [{\citenamefont {Perkins}\ and\ \citenamefont
  {Xu}(2021)}]{perkinsFrozen1RSBStructure2021}%
  \BibitemOpen
  \bibfield  {author} {\bibinfo {author} {\bibfnamefont {W.}~\bibnamefont
  {Perkins}}\ and\ \bibinfo {author} {\bibfnamefont {C.}~\bibnamefont {Xu}},\
  }in\ \href {\doibase 10.1145/3406325.3451119} {\emph {\bibinfo {booktitle}
  {Proceedings of the 53rd {{Annual ACM SIGACT Symposium}} on {{Theory}} of
  {{Computing}}}}},\ \bibinfo {series and number} {{{STOC}} 2021}\ (\bibinfo
  {publisher} {{Association for Computing Machinery}},\ \bibinfo {address}
  {{New York, NY, USA}},\ \bibinfo {year} {2021})\ pp.\ \bibinfo {pages}
  {1579--1588}\BibitemShut {NoStop}%
\bibitem [{\citenamefont {Abbe}\ \emph
  {et~al.}(2021{\natexlab{a}})\citenamefont {Abbe}, \citenamefont {Li},\ and\
  \citenamefont {Sly}}]{abbeProofContiguityConjecture2021}%
  \BibitemOpen
  \bibfield  {author} {\bibinfo {author} {\bibfnamefont {E.}~\bibnamefont
  {Abbe}}, \bibinfo {author} {\bibfnamefont {S.}~\bibnamefont {Li}}, \ and\
  \bibinfo {author} {\bibfnamefont {A.}~\bibnamefont {Sly}},\ }\href@noop {} {\
   (\bibinfo {year} {2021}{\natexlab{a}})},\ \Eprint
  {http://arxiv.org/abs/2102.13069} {arXiv:2102.13069 [math-ph, stat]}
  \BibitemShut {NoStop}%
\bibitem [{\citenamefont {Braunstein}\ and\ \citenamefont
  {Zecchina}(2006)}]{braunsteinLearningMessagePassing2006}%
  \BibitemOpen
  \bibfield  {author} {\bibinfo {author} {\bibfnamefont {A.}~\bibnamefont
  {Braunstein}}\ and\ \bibinfo {author} {\bibfnamefont {R.}~\bibnamefont
  {Zecchina}},\ }\href {\doibase 10.1103/PhysRevLett.96.030201} {\bibfield
  {journal} {\bibinfo  {journal} {Physical Review Letters}\ }\textbf {\bibinfo
  {volume} {96}},\ \bibinfo {pages} {030201} (\bibinfo {year}
  {2006})}\BibitemShut {NoStop}%
\bibitem [{\citenamefont {Baldassi}\ \emph
  {et~al.}(2016{\natexlab{b}})\citenamefont {Baldassi}, \citenamefont {Borgs},
  \citenamefont {Chayes}, \citenamefont {Ingrosso}, \citenamefont {Lucibello},
  \citenamefont {Saglietti},\ and\ \citenamefont
  {Zecchina}}]{baldassiUnreasonableEffectivenessLearning2016}%
  \BibitemOpen
  \bibfield  {author} {\bibinfo {author} {\bibfnamefont {C.}~\bibnamefont
  {Baldassi}}, \bibinfo {author} {\bibfnamefont {C.}~\bibnamefont {Borgs}},
  \bibinfo {author} {\bibfnamefont {J.}~\bibnamefont {Chayes}}, \bibinfo
  {author} {\bibfnamefont {A.}~\bibnamefont {Ingrosso}}, \bibinfo {author}
  {\bibfnamefont {C.}~\bibnamefont {Lucibello}}, \bibinfo {author}
  {\bibfnamefont {L.}~\bibnamefont {Saglietti}}, \ and\ \bibinfo {author}
  {\bibfnamefont {R.}~\bibnamefont {Zecchina}},\ }\href {\doibase
  10.1073/pnas.1608103113} {\bibfield  {journal} {\bibinfo  {journal}
  {Proceedings of the National Academy of Sciences}\ }\textbf {\bibinfo
  {volume} {113}},\ \bibinfo {pages} {E7655} (\bibinfo {year}
  {2016}{\natexlab{b}})},\ \Eprint {http://arxiv.org/abs/1605.06444}
  {arXiv:1605.06444} \BibitemShut {NoStop}%
\bibitem [{\citenamefont {Abbe}\ \emph
  {et~al.}(2021{\natexlab{b}})\citenamefont {Abbe}, \citenamefont {Li},\ and\
  \citenamefont {Sly}}]{abbeBinaryPerceptronEfficient2021}%
  \BibitemOpen
  \bibfield  {author} {\bibinfo {author} {\bibfnamefont {E.}~\bibnamefont
  {Abbe}}, \bibinfo {author} {\bibfnamefont {S.}~\bibnamefont {Li}}, \ and\
  \bibinfo {author} {\bibfnamefont {A.}~\bibnamefont {Sly}},\ }\href@noop {} {\
   (\bibinfo {year} {2021}{\natexlab{b}})},\ \Eprint
  {http://arxiv.org/abs/2111.03084} {arXiv:2111.03084 [math-ph, stat]}
  \BibitemShut {NoStop}%
\bibitem [{\citenamefont {Baldassi}\ \emph {et~al.}(2021)\citenamefont
  {Baldassi}, \citenamefont {Lauditi}, \citenamefont {Malatesta}, \citenamefont
  {Perugini},\ and\ \citenamefont
  {Zecchina}}]{baldassiUnveilingStructureWide2021}%
  \BibitemOpen
  \bibfield  {author} {\bibinfo {author} {\bibfnamefont {C.}~\bibnamefont
  {Baldassi}}, \bibinfo {author} {\bibfnamefont {C.}~\bibnamefont {Lauditi}},
  \bibinfo {author} {\bibfnamefont {E.~M.}\ \bibnamefont {Malatesta}}, \bibinfo
  {author} {\bibfnamefont {G.}~\bibnamefont {Perugini}}, \ and\ \bibinfo
  {author} {\bibfnamefont {R.}~\bibnamefont {Zecchina}},\ }\href {\doibase
  10.1103/PhysRevLett.127.278301} {\bibfield  {journal} {\bibinfo  {journal}
  {Physical Review Letters}\ }\textbf {\bibinfo {volume} {127}},\ \bibinfo
  {pages} {278301} (\bibinfo {year} {2021})},\ \bibinfo {note} {publisher:
  American Physical Society}\BibitemShut {NoStop}%
\bibitem [{\citenamefont {Cavaliere}\ \emph {et~al.}(2021)\citenamefont
  {Cavaliere}, \citenamefont {Lesieur},\ and\ \citenamefont
  {{Ricci-Tersenghi}}}]{cavaliereOptimizationDynamicTransition2021}%
  \BibitemOpen
  \bibfield  {author} {\bibinfo {author} {\bibfnamefont {A.~G.}\ \bibnamefont
  {Cavaliere}}, \bibinfo {author} {\bibfnamefont {T.}~\bibnamefont {Lesieur}},
  \ and\ \bibinfo {author} {\bibfnamefont {F.}~\bibnamefont
  {{Ricci-Tersenghi}}},\ }\href {\doibase 10.1088/1742-5468/ac382e} {\bibfield
  {journal} {\bibinfo  {journal} {Journal of Statistical Mechanics: Theory and
  Experiment}\ }\textbf {\bibinfo {volume} {2021}},\ \bibinfo {pages} {113302}
  (\bibinfo {year} {2021})}\BibitemShut {NoStop}%
\bibitem [{\citenamefont {Yoshino}(2018)}]{yoshinoDisorderfreeSpinGlass2018}%
  \BibitemOpen
  \bibfield  {author} {\bibinfo {author} {\bibfnamefont {H.}~\bibnamefont
  {Yoshino}},\ }\href {\doibase 10.21468/SciPostPhys.4.6.040} {\bibfield
  {journal} {\bibinfo  {journal} {SciPost Physics}\ }\textbf {\bibinfo {volume}
  {4}},\ \bibinfo {pages} {040} (\bibinfo {year} {2018})}\BibitemShut {NoStop}%
\bibitem [{\citenamefont {Mari}\ \emph {et~al.}(2009)\citenamefont {Mari},
  \citenamefont {Krzakala},\ and\ \citenamefont
  {Kurchan}}]{mariJammingGlassTransitions2009}%
  \BibitemOpen
  \bibfield  {author} {\bibinfo {author} {\bibfnamefont {R.}~\bibnamefont
  {Mari}}, \bibinfo {author} {\bibfnamefont {F.}~\bibnamefont {Krzakala}}, \
  and\ \bibinfo {author} {\bibfnamefont {J.}~\bibnamefont {Kurchan}},\ }\href
  {\doibase 10.1103/PhysRevLett.103.025701} {\bibfield  {journal} {\bibinfo
  {journal} {Physical Review Letters}\ }\textbf {\bibinfo {volume} {103}},\
  \bibinfo {pages} {025701} (\bibinfo {year} {2009})}\BibitemShut {NoStop}%
\bibitem [{\citenamefont {M{\'e}zard}\ \emph {et~al.}(2011)\citenamefont
  {M{\'e}zard}, \citenamefont {Parisi}, \citenamefont {Tarzia},\ and\
  \citenamefont {Zamponi}}]{mezardSolutionSolvableModel2011}%
  \BibitemOpen
  \bibfield  {author} {\bibinfo {author} {\bibfnamefont {M.}~\bibnamefont
  {M{\'e}zard}}, \bibinfo {author} {\bibfnamefont {G.}~\bibnamefont {Parisi}},
  \bibinfo {author} {\bibfnamefont {M.}~\bibnamefont {Tarzia}}, \ and\ \bibinfo
  {author} {\bibfnamefont {F.}~\bibnamefont {Zamponi}},\ }\href {\doibase
  10.1088/1742-5468/2011/03/P03002} {\bibfield  {journal} {\bibinfo  {journal}
  {Journal of Statistical Mechanics: Theory and Experiment}\ }\textbf {\bibinfo
  {volume} {2011}},\ \bibinfo {pages} {P03002} (\bibinfo {year}
  {2011})}\BibitemShut {NoStop}%
\bibitem [{\citenamefont {Dawson}\ \emph {et~al.}(2000)\citenamefont {Dawson},
  \citenamefont {Foffi}, \citenamefont {Fuchs}, \citenamefont {G{\"o}tze},
  \citenamefont {Sciortino}, \citenamefont {Sperl}, \citenamefont {Tartaglia},
  \citenamefont {Voigtmann},\ and\ \citenamefont
  {Zaccarelli}}]{dawsonHigherorderGlasstransitionSingularities2000}%
  \BibitemOpen
  \bibfield  {author} {\bibinfo {author} {\bibfnamefont {K.}~\bibnamefont
  {Dawson}}, \bibinfo {author} {\bibfnamefont {G.}~\bibnamefont {Foffi}},
  \bibinfo {author} {\bibfnamefont {M.}~\bibnamefont {Fuchs}}, \bibinfo
  {author} {\bibfnamefont {W.}~\bibnamefont {G{\"o}tze}}, \bibinfo {author}
  {\bibfnamefont {F.}~\bibnamefont {Sciortino}}, \bibinfo {author}
  {\bibfnamefont {M.}~\bibnamefont {Sperl}}, \bibinfo {author} {\bibfnamefont
  {P.}~\bibnamefont {Tartaglia}}, \bibinfo {author} {\bibfnamefont
  {T.}~\bibnamefont {Voigtmann}}, \ and\ \bibinfo {author} {\bibfnamefont
  {E.}~\bibnamefont {Zaccarelli}},\ }\href {\doibase
  10.1103/PhysRevE.63.011401} {\bibfield  {journal} {\bibinfo  {journal}
  {Physical Review E}\ }\textbf {\bibinfo {volume} {63}},\ \bibinfo {pages}
  {011401} (\bibinfo {year} {2000})}\BibitemShut {NoStop}%
\bibitem [{\citenamefont {Sciortino}(2002)}]{sciortinoOneLiquidTwo2002}%
  \BibitemOpen
  \bibfield  {author} {\bibinfo {author} {\bibfnamefont {F.}~\bibnamefont
  {Sciortino}},\ }\href {\doibase 10.1038/nmat752} {\bibfield  {journal}
  {\bibinfo  {journal} {Nature Materials}\ }\textbf {\bibinfo {volume} {1}},\
  \bibinfo {pages} {145} (\bibinfo {year} {2002})}\BibitemShut {NoStop}%
\bibitem [{\citenamefont {Parisi}\ \emph {et~al.}(2020)\citenamefont {Parisi},
  \citenamefont {Urbani},\ and\ \citenamefont
  {Zamponi}}]{parisiTheorySimpleGlasses2020}%
  \BibitemOpen
  \bibfield  {author} {\bibinfo {author} {\bibfnamefont {G.}~\bibnamefont
  {Parisi}}, \bibinfo {author} {\bibfnamefont {P.}~\bibnamefont {Urbani}}, \
  and\ \bibinfo {author} {\bibfnamefont {F.}~\bibnamefont {Zamponi}},\ }\href
  {\doibase 10.1017/9781108120494} {\emph {\bibinfo {title} {Theory of {{Simple
  Glasses}}: {{Exact Solutions}} in {{Infinite Dimensions}}}}}\ (\bibinfo
  {publisher} {{Cambridge University Press}},\ \bibinfo {address}
  {{Cambridge}},\ \bibinfo {year} {2020})\BibitemShut {NoStop}%
\bibitem [{\citenamefont {Mari}\ and\ \citenamefont
  {Kurchan}(2011)}]{mariDynamicalTransitionGlasses2011}%
  \BibitemOpen
  \bibfield  {author} {\bibinfo {author} {\bibfnamefont {R.}~\bibnamefont
  {Mari}}\ and\ \bibinfo {author} {\bibfnamefont {J.}~\bibnamefont {Kurchan}},\
  }\href {\doibase 10.1063/1.3626802} {\bibfield  {journal} {\bibinfo
  {journal} {The Journal of Chemical Physics}\ }\textbf {\bibinfo {volume}
  {135}},\ \bibinfo {pages} {124504} (\bibinfo {year} {2011})},\ \Eprint
  {http://arxiv.org/abs/1104.3420} {arXiv:1104.3420} \BibitemShut {NoStop}%
\bibitem [{\citenamefont {Krzakala}\ and\ \citenamefont
  {Zdeborov{\'a}}(2009)}]{krzakalaHidingQuietSolutions2009}%
  \BibitemOpen
  \bibfield  {author} {\bibinfo {author} {\bibfnamefont {F.}~\bibnamefont
  {Krzakala}}\ and\ \bibinfo {author} {\bibfnamefont {L.}~\bibnamefont
  {Zdeborov{\'a}}},\ }\href {\doibase 10.1103/PhysRevLett.102.238701}
  {\bibfield  {journal} {\bibinfo  {journal} {Physical Review Letters}\
  }\textbf {\bibinfo {volume} {102}},\ \bibinfo {pages} {238701} (\bibinfo
  {year} {2009})}\BibitemShut {NoStop}%
\bibitem [{\citenamefont {Gabri\'e}\ \emph {et~al.}(2023)\citenamefont
  {Gabri\'e}, \citenamefont {Ganguli}, \citenamefont {Lucibello},\ and\
  \citenamefont {Zecchina}}]{GabrieFarBeyond}%
  \BibitemOpen
  \bibfield  {author} {\bibinfo {author} {\bibfnamefont {M.}~\bibnamefont
  {Gabri\'e}}, \bibinfo {author} {\bibfnamefont {S.}~\bibnamefont {Ganguli}},
  \bibinfo {author} {\bibfnamefont {C.}~\bibnamefont {Lucibello}}, \ and\
  \bibinfo {author} {\bibfnamefont {R.}~\bibnamefont {Zecchina}},\ }\enquote
  {\bibinfo {title} {Neural networks: From the perceptron to deep nets},}\ in\
  \href@noop {} {\emph {\bibinfo {booktitle} {Spin Glass Theory and Far Beyond
  --- Replica Symmetry Breaking after 40 Years}}},\ \bibinfo {editor} {edited
  by\ \bibinfo {editor} {\bibfnamefont {P.}~\bibnamefont {Charbonneau}},
  \bibinfo {editor} {\bibfnamefont {M.}~\bibnamefont {M\'ezard}}, \bibinfo
  {editor} {\bibfnamefont {E.}~\bibnamefont {Marinari}}, \bibinfo {editor}
  {\bibfnamefont {F.}~\bibnamefont {Ricci-Tersenghi}}, \bibinfo {editor}
  {\bibfnamefont {G.}~\bibnamefont {Sicuro}}, \ and\ \bibinfo {editor}
  {\bibfnamefont {F.}~\bibnamefont {Zamponi}}}\ (\bibinfo  {publisher} {World
  Scientific},\ \bibinfo {address} {Singapore},\ \bibinfo {year} {2023})\
  Chap.~\bibinfo {chapter} {24}, pp.\ \bibinfo {pages} {477--497}\BibitemShut
  {NoStop}%
\bibitem [{\citenamefont {Angelini}\ and\ \citenamefont
  {Ricci-Tersenghi}(2022)}]{angeliniLimits2022}%
  \BibitemOpen
  \bibfield  {author} {\bibinfo {author} {\bibfnamefont {M.~C.}\ \bibnamefont
  {Angelini}}\ and\ \bibinfo {author} {\bibfnamefont {F.}~\bibnamefont
  {Ricci-Tersenghi}},\ }\href {\doibase 10.48550/arXiv.2206.04760} {\enquote
  {\bibinfo {title} {A theory explaining the limits and performances of
  algorithms based on simulated annealing in solving sparse hard inference
  problems},}\ } (\bibinfo {year} {2022}),\ \bibinfo {note}
  {arXiv:2206.04760}\BibitemShut {NoStop}%
\end{thebibliography}%

\end{document}